\def\aa{{A\&A}}
\def\aas{{A\&AS}}
\def\aj{{AJ}}

\def\mnras{{MNRAS}}

\def\pasp{{PASP}}

\documentclass[cup5b]{caps}
\usepackage{graphicx}
\usepackage{amssymb}
\usepackage{ociwsymp4e}  

\HeadText{Allende Prieto et al.}  

\begin{document}

\pagenumbering{arabic}


%
\author[]{C. ALLENDE PRIETO$^{1}$, T. C. BEERS$^{2}$, 
 Y. LI$^{2}$,  H. J. NEWBERG$^{3}$, R. WILHELM$^{4}$ 
and B. YANNY$^{5}$ 
\\
(1) University of Texas, Austin, TX, USA\\
(2) Michigan State University, East Lansing, MI, USA\\
(3) Rensselaer Polytechnical Institute, Troy, NY, USA\\
(4) Texas Tech University, Lubbock, TX, USA\\
(5) Fermi National Accelerator Laboratory, Batavia, IL, USA }

\chapter{New Resources to Explore the \\  Old Galaxy: Mining the SDSS }

\begin{abstract}

The Sloan Digital Sky Survey (SDSS) is collecting photometry and intermediate
resolution spectra for 
$\sim 10^5$ stars in the thick-disk and stellar halo of the Milky Way. This
massive dataset can be used to infer the properties of the stars that make up
these structures, and considerably deepen our vision of the old components 
of the Galaxy. We devise tools for automatic analysis of the SDSS photometric
and spectroscopic data based on plane-parallel line-blanketed LTE 
model atmospheres
and fast optimization algorithms. A preliminary study of about 5000 stars in 
the Early Data Release gives a hint of the vast amount of information that
the SDSS stellar sample contains.

\end{abstract}

\section{Introduction}

The Sloan Digital Sky Survey is an ambitious project that is imaging
about one fourth of the sky with five broad-band filters. 
The survey includes followup 
intermediate-dispersion ($\lambda/\delta\lambda \simeq 1800$) 
spectroscopy (York et al. 2000).
The final catalog is expected to include photometry and spectroscopy 
for about $10^8$ and $10^6$ sources, respectively. Focused on  
extragalactic science, the spectroscopic survey aims at amassing the 
largest possible collection of galaxy redshifts. The dedicated f/5
Ritchey-Chr\'etien-like 2.5m telescope has a three-degree field of view. 
In the spectroscopic mode, up to 640 (180$\mu$m or 3 arcsec \O) 
fibers can be simultaneously positioned on the focal plane to feed two
identical spectrographs. Each spectrograph has a blue and a red arm that 
provide continuous coverage in the range 381--910 nm. 

The selection criteria for the spectroscopic targets are rather complex
(Eisenstein et al. 2001; Richards et al. 2002; Stoughton et al. 2002; 
Strauss et al. 2002).
Galaxies and quasar candidates take about 90\% of the fibers, with the
remaining used to observe the sky background and Galactic stars, which 
are either selected  for being 
 {\it peculiar} (brown dwarfs, blue-horizontal branch stars, 
carbon stars, etc.), or intended for reddening and flux calibration. 
In addition, almost a third of the quasar candidates in 
the Early Data Release  turned out to be stars. Nearly $\sim 10^5$ stellar
spectra will be released by the end of the survey in 2006. With exposure 
times per plate of the order of 45 minutes, the targeted stars have
$V$ magnitudes in the range 14--21, signal-to-noise ratios ($S/N$) 
between 5 and 150, and lie at distances of 
up to hundreds of kiloparsecs from the Galactic plane. 
When released, the SDSS stellar spectra will constitute 
the largest spectroscopic survey  
of the Galactic thick-disk and halo populations yet assembled.

\section{Analysis}

The analysis of a massive dataset calls for 
automated procedures. The SDSS images and spectra are processed
by a series of pipelines specialized in tasks such as astrometric, wavelength, 
and flux calibrations, aperture photometry, or the extraction and 
classification of spectra. Starting from the released photometry and
spectra for Galactic stars, we are interested in a detailed classification
based on the fundamental atmospheric parameters. Projected radial velocities
are directly measurable from the spectra. Given the large spectral
coverage, we expect to be able to quantify the 
interstellar reddening towards the observed stars. Stellar 
 metallicities, and even perhaps the abundance ratios of chemical elements 
 or groups of elements well-represented 
 in the  spectra, 
are obviously some of the most valuable
information to search for. It is also most 
interesting to use the derived stellar parameters, chemical abundances,
 and interstellar reddening to infer distances and ages.

We make use of
 the SDSS ($ugriz$) photometry and the $\sim 3800$ pixels in an
 object's spectrum altogether. 
 We found it helpful to trade resolution for $S/N$, and therefore
the spectra are smoothed to $\lambda/\delta\lambda = 1000$ 
by convolution with a Gaussian profile.
 As absolute fluxes are not relevant at this point, we use
photometric indices and normalize the spectra $\{S_i\}$ to satisfy
$\sum_{i=1}^{m} S_i/m = 0.5$. 
The relevant data vector is 
${\bf T} \equiv \{u-g,g-r,r-i,i-z,S_1,S_2,S_3,\dots, S_m\}$, where $m=2600$. 
We model {\bf T} with plane-parallel line-blanketed LTE model atmospheres
and radiative transfer calculations, 
as a function of the stellar parameters (effective
temperature $T_{\rm eff}$, surface gravity $g$, 
and overall metallicity 
[Fe/H]\footnote{[E/H] $= \log \frac{N({\rm E})}{N({\rm H})} - 
\log \left(\frac{N({\rm E})}{N({\rm H})}\right)_{\odot}$, 
where $N$ represents number density of
a chemical element.}). The collection of model atmospheres and
low-resolution synthetic spectra of Kurucz (1993) is used. This grid
was calculated with a mixing-length $l/H_p = 1.25$, and a micro-turbulence
of 2 km s$^{-1}$. The low-dispersion
spectra are convolved with the SDSS filter responses (Strauss \& Gunn 2001).
The atmospheric structures are used to produce LTE synthetic spectra with a
resolving power $\lambda/\delta\lambda = 1000$ between 381 and 910 nm.
Balmer line profiles are treated as in Hubeny, Hummer, \& Lanz (1994). 
The radiative transfer equation is solved with the code {\tt synspec}
(Hubeny \& Lanz 2000), using very simple continuous opacities: H, H$^{-}$,
Rayleigh and electron scattering (with the prescriptions in Hubeny 1988).
The calculations included 131821 atomic line transitions, but no molecular
features. 

Both photometric magnitudes and spectra are computed for a discreet
$12\times4\times6$ grid  spanning the ranges 4500 to  10000 K, 2.0 to 5.0 dex, 
and $-4.5$ to $+0.5$ dex in $T_{\rm eff}$, $\log g$ (c.g.s units), 
and [Fe/H], respectively.
The interstellar reddening ($E(B-V)$)
is parameterized as in Fitzpatrick (1999),
 adopting $R \equiv A(V)/E(B-V) = 3.1$. This parameter gives one more dimension
 to the grid. We consider $E(B-V)$ in the range $0.0-0.1$ with  just 
 three values. Model spectra and photometry for sets of parameters
 off the grid nodes are derived by multi-linear interpolation.

Some elements show abundance ratios to iron that are non-solar in metal-poor
stars. This is largely ignored in our modeling. However,  we 
consider  enhancements to the abundances of Mg and Ca in metal-poor stars
when calculating synthetic spectra because these elements produce
 strong lines on which our analysis heavily relies. Following
Beers et al. (1999) we adopt [$\alpha$/Fe] $\simeq$ [Ca/Fe] $\simeq$ [Mg/Fe]:

\begin{equation}
[\alpha/Fe] = \left\{ 
\begin{tabular}{cc}
  		           0 		& if $ 0 \ge$ [Fe/H] 	\\
  		      
        	      $-0.267$ [Fe/H] 	& if $-1.5 \le$ [Fe/H] $< 0$ \\
 	
  		            $+0.4$ 	& if  [Fe/H] $< -1.5$.   \\
\end{tabular}
\right.
\label{law}
\end{equation}

We perform a search for the  model parameters that minimize the 
distance between the model {\bf T} ($T_{\rm eff}, g$, [Fe/H], $E(B-V)$) and the 
 observations' vector {\bf O}. In a $\chi^2$ fashion, we define such distance as
 
 \begin{equation}
 \mu = \sum_{i=1}^{m+4}{W_i (O_i - T_i)^2}
 \label{fit}
 \end{equation}
 
 \noindent where the weights $W_i$ are optimized for performance.
 The search is accomplished by using either the Nelder-Mead simplex method
 (Nelder \& Mead 1965) or a genetic algorithm (Carroll 1999). 
The multi-linear interpolation  was deemed as accurate enough after
repeating the  interpolations with the decimal logarithm of the 
fluxes. Testing with finer grids in
 $\log g$, [Fe/H], and $E(B-V)$, 
 led only to marginal variations in the results. Classification of a single
 spectrum takes a few seconds on a 600 MHz workstation.

   \begin{figure}
    \centering
    \includegraphics[width=8cm,angle=90]{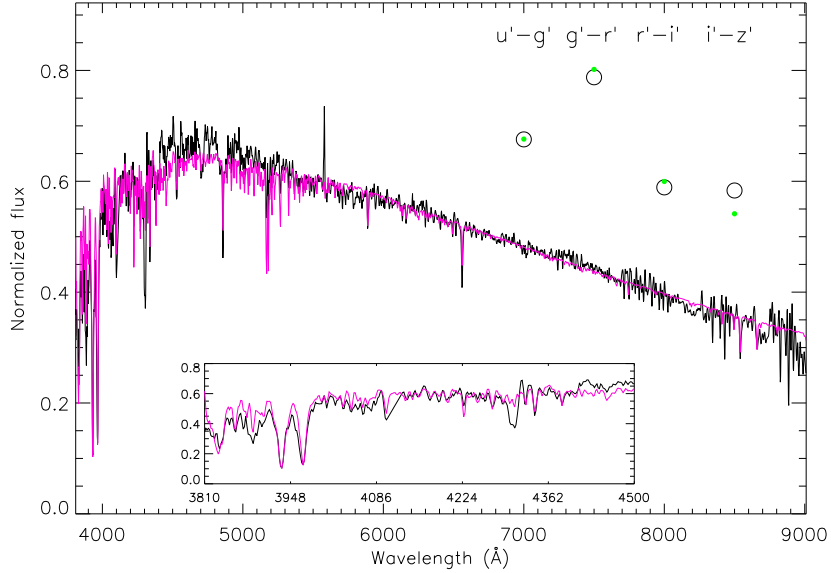}
    \caption{Comparison of observed spectra (black line) and photometric 
    indices (open circles) with the best-fitting model (red line and green
    circles) for one of the EDR stars. The inset shows an enlarge view of
    the bluest part of the spectrum. The primes in the photometric magnitudes
    are related to technical subtleties between different 
    calibrations. For our purposes these are the same as the
    non-primed magnitudes we refer to in the text.}
    \label{example}
  \end{figure}

 Fig. \ref{example} shows an example of observed (black) and model 
 (red and green) 
 fluxes for
 one of the EDR stars that we fit with $T_{\rm eff} = 5388$ K, $\log g = 4.71$,
 [Fe/H]$= -1.2$, and $E(B-V)= 0.004$.  The inset plot gives an expanded view
 of the blue part of the spectrum. 
 Note that molecular features, such as the G band (CH) at $\sim 4300$ \AA\
 are not reproduced by the model spectrum.  
 The photometric indices have been
 scaled to fit in the graph's box and placed at arbitrary wavelengths.

 Once the atmospheric parameters are defined, we make use
 of stellar evolutionary calculations by the Padova group 
 (Alongi et al. 1993; Bressan et al. 1993; Fagotto et al. 1994; Bertelli
 et al. 1994) to find the best estimates for other stellar parameters:
 radius, $M_V$, mass ($M$), $Age$, etc. With the atmospheric 
 parameters and their 
 uncertainties  in hand
 we define a normalized probability density distribution that is Normal
 for $T_{\rm eff}$ and $\log g$, and a boxcar function in $\log (Z/Z_{\odot})$ 
 
 \begin{equation}
P  \propto
\exp \left[ -\left(\frac{T_{\rm eff}-T_{\rm eff}^{*}}{\sqrt{2} \sigma(T_{\rm eff})}\right)^2\right]
\exp \left[ -\left(\frac{\log g - \log g^{*}}{\sqrt{2} \sigma(\log g)}\right)^2\right]
B\left(\log (Z/Z_{\odot})\right),
\label{triplegauss}
\end{equation}
 
 \noindent  which is then used to find the best estimate 
 of a stellar parameter $X$
 by integration over the space 
 ($Z/H$, Age, and initial mass $M$) that
 characterizes the stellar isochrones of Bertelli et al. (1994)
 
\begin{equation}
\bar{X} = \int_{Z/H} \int_{Age} \int_{M} X P(Z/H,Age,M) d(Z/Z_{\odot}) d(Age) dM.
\end{equation} 
 
The isochrones employed do not consider
enhancements in the abundances of the $\alpha$ elements for metal-poor stars.
Thus, we simply equate [Fe/H] $ = \log (Z/Z_{\odot})$.
More realistic relations should take this into account, 
and will be explored in the future. 

Finally, the magnitudes in the SDSS passbands can be used to estimate the 
Johnson $V$ magnitudes of the stars (Zhao \& Newberg 2002). Knowing $M_V$ 
and the reddening, it is then straightforward to derive distances. 

 \begin{figure}
    \centering
    \includegraphics[width=8cm,angle=0]{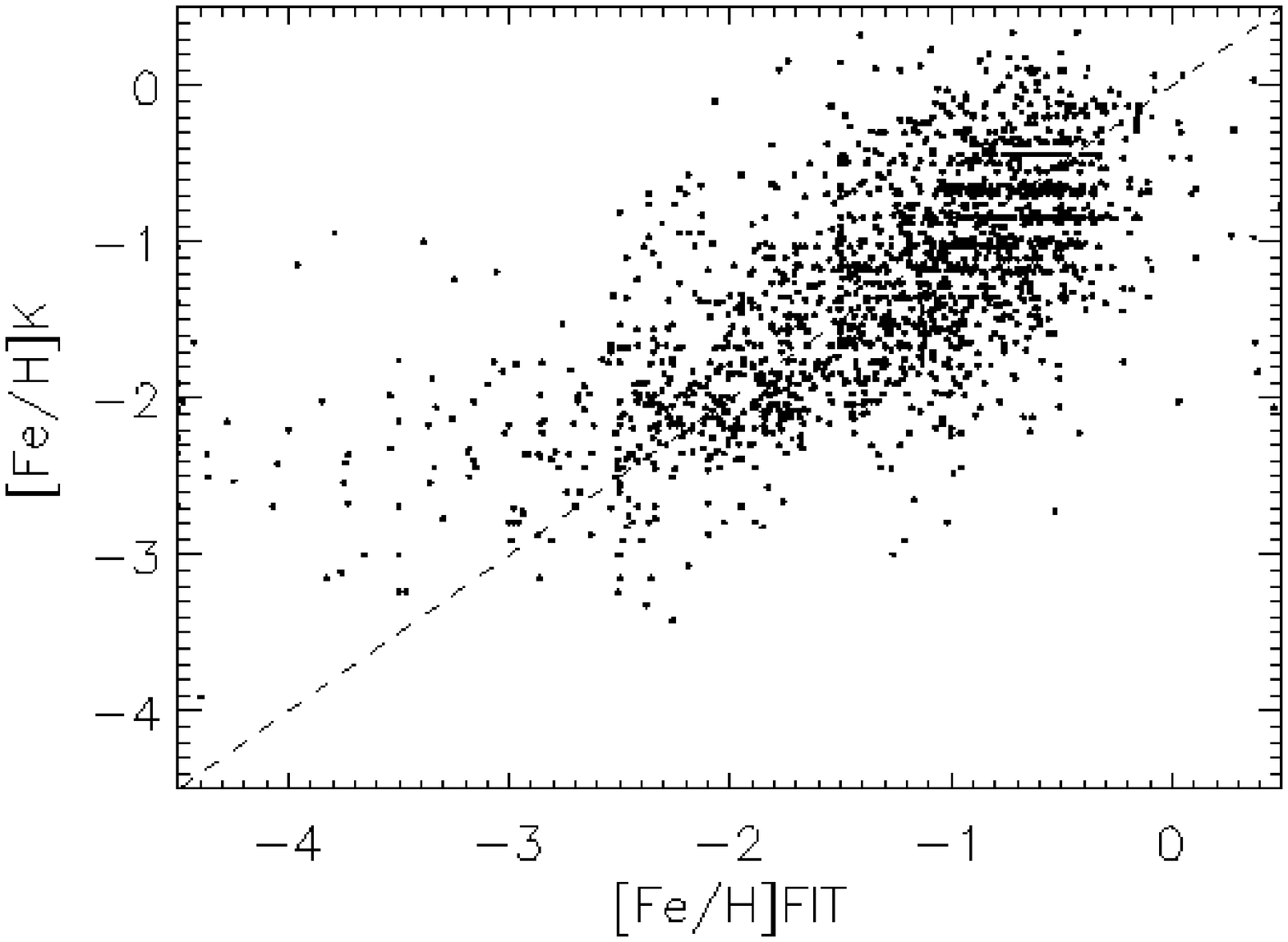}
     \includegraphics[width=8cm,angle=0]{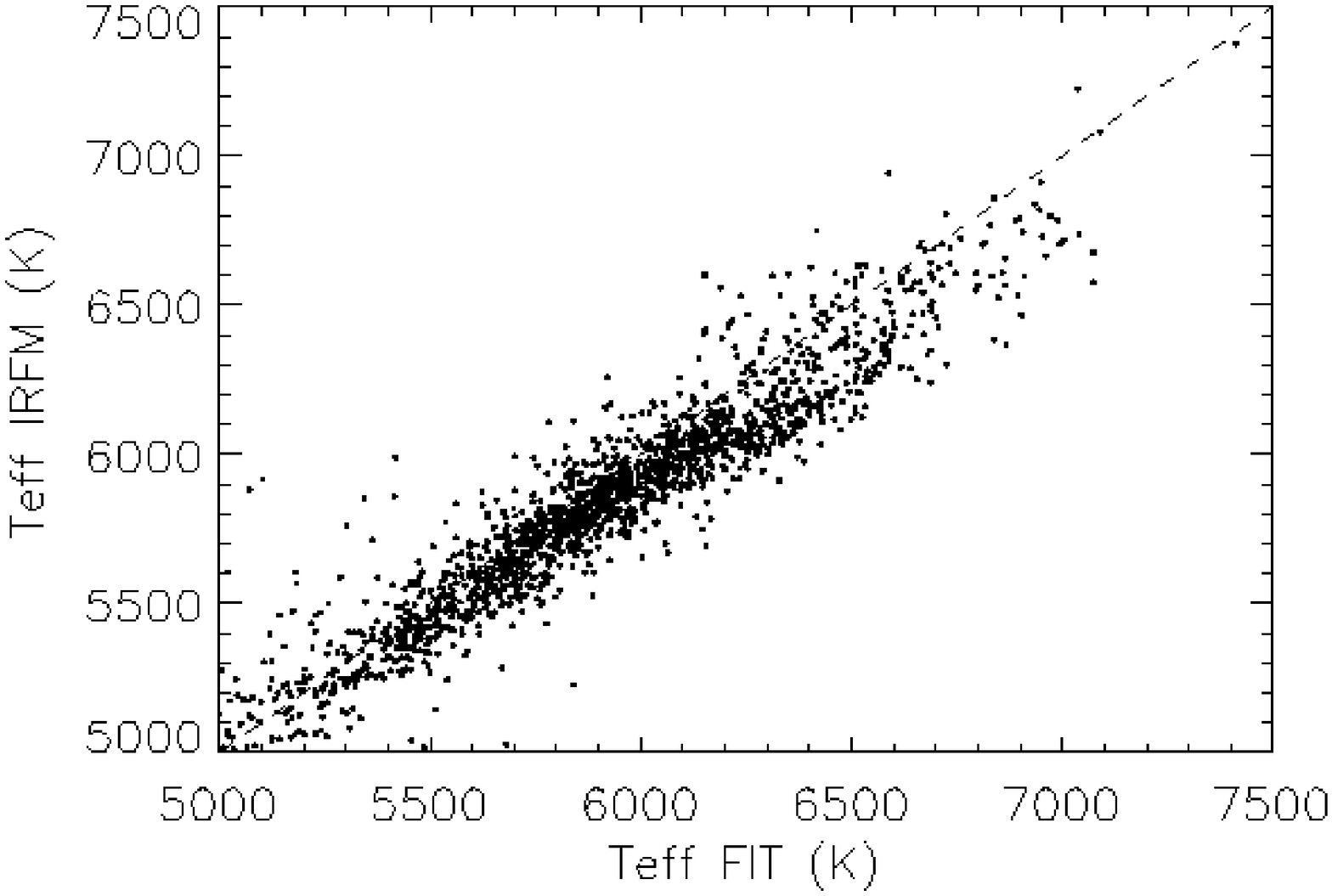}
    \caption{Top panel: Comparison between the [Fe/H] derived by our spectral 
    fitting technique and those with the K method of Beers et al. (1999). Lower
    panel: The effective temperatures from fitting the spectra are compared to 
    those based on the photometry and the IRFM calibrations of Alonso et al.
    (1996, 1999). The dashed lines have a slope of one.}
    \label{k}
  \end{figure}

\section{Checking on $T_{\rm eff}$ and [Fe/H]}
 
The inclusion of the Ca K line and the seven first members of the Balmer 
series in the EDR spectra 
makes them suitable
for the application of  well-tested techniques 
developed for the followup of  stars in the HK survey 
(Beers, Preston \& Shectman 1992; Beers et al. 1990;  Beers et al. 1999). 

After measuring the pseudo-equivalent widths of the relevant spectral
features and estimating the $(B-V)$ colors from the SDSS photometry, 
we obtain a second, relatively independent, measure of 
 the metallicities of the EDR stars.
After visual inspection, a subsample of 1910 stars was
deemed of reasonably good quality, and their metallicities  are compared
with those determined by spectral fitting in the upper panel of Fig. \ref{k}.
The overall agreement for [Fe/H] $> -2.5$ is reasonable, with a scatter of
about 0.5 dex. A systematic discrepancy is apparent for the most metal-poor
stars. This disagreement exceeds the internal uncertainties of each method
and should be investigated. We have found that, in the presence of severe
noise, the optimization algorithms tend to underestimate the metallicity. 
We should also note that a spurious feature is apparent
in  many EDR spectra right between the Ca H and K lines, which might be
affecting the Ca II K method.

The $(B-V)$ colors estimated from the SDSS $ugriz$  photometry and the stellar
metallicities were fed to the photometric calibrations of
Alonso et al. (1996, 1999), which are based on the Infrared Flux Method
(IRFM; Blackwell, Shallis \& Selby 1979). 
These  calibrations have an internal scatter
of about 2\% and an uncertainty in the zero point of about 1\%. 
Therefore, they offer a reliable external check to the $T_{\rm eff}$s 
determined automatically from the simultaneous analysis of  EDR 
spectra and 
photometry. The interstellar reddening was corrected
using the values determined spectroscopically. The lower panel of Fig. \ref{k}
shows a pleasing correspondence between the two $T_{\rm eff}$ determinations. 
The IRFM $T_{\rm eff}$s are, on average, lower by 66 K with an rms
scatter between the two scales of 160 K.

\section{Application to the EDR. Preliminary results.}

\begin{figure}[t!]
    \centering
    \includegraphics[width=5cm,angle=0]{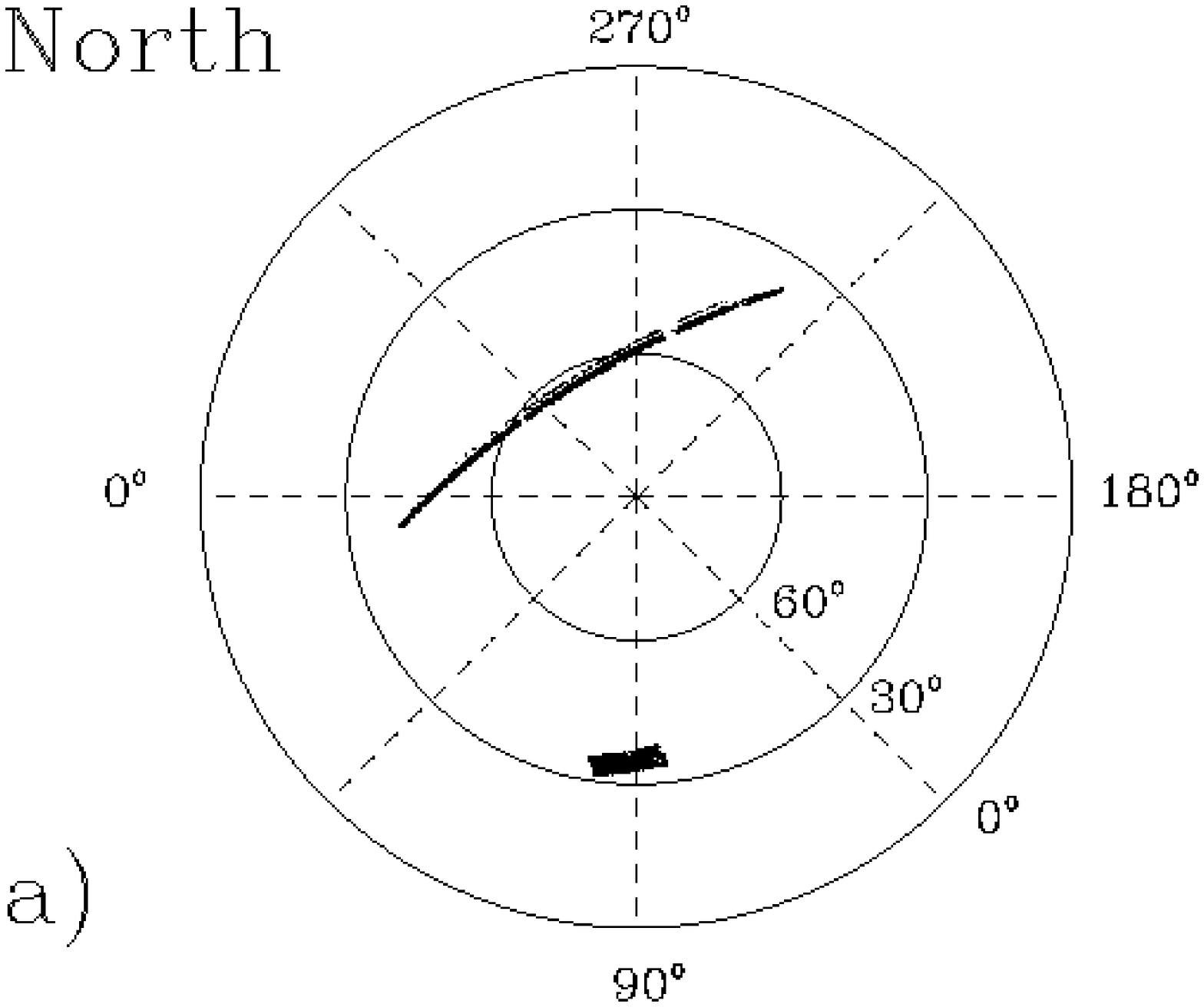}
    \includegraphics[width=5cm,angle=0]{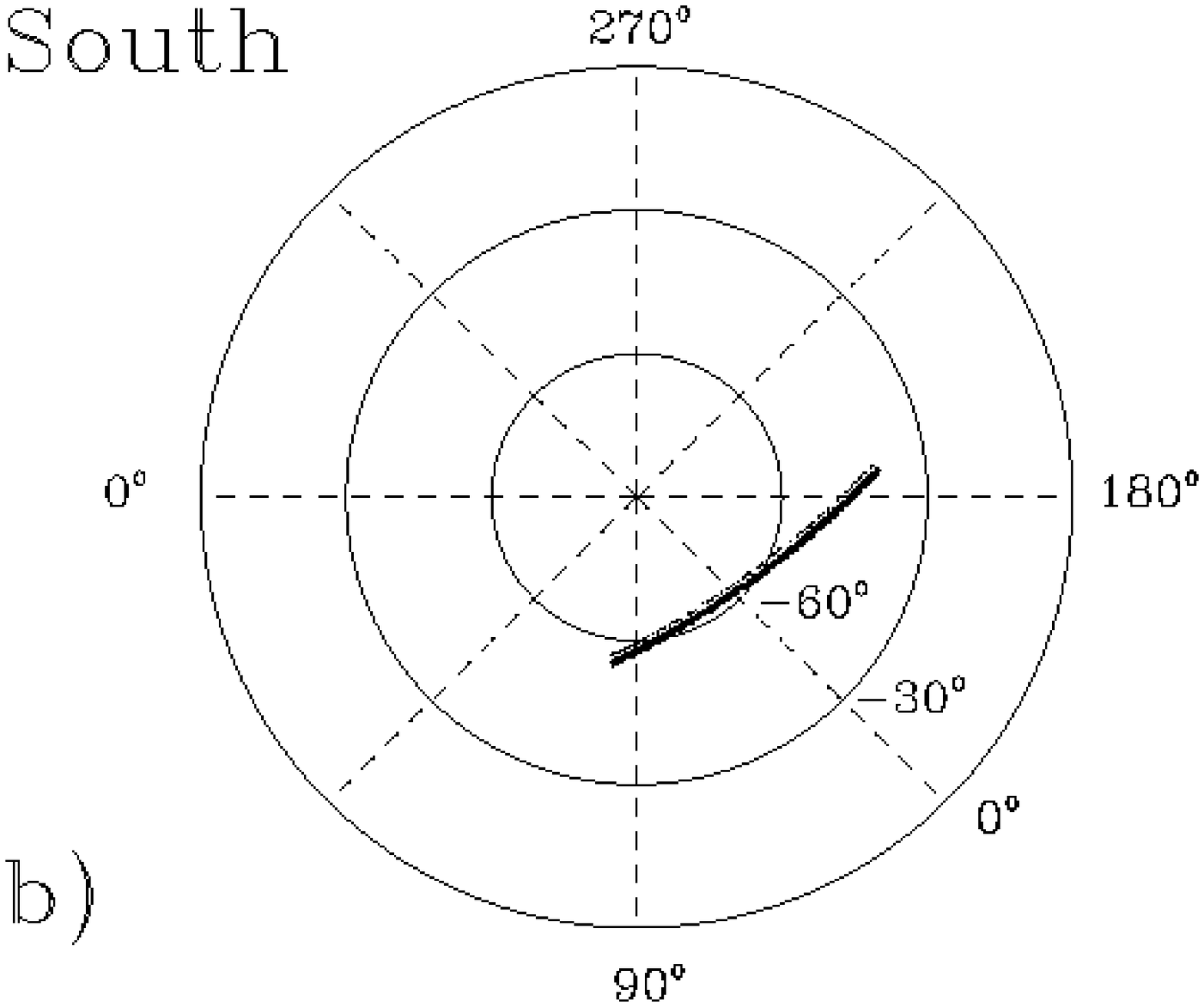}
    \caption{Location of the EDR stars in galactic coordinates for each
     hemisphere. The galactic
    latitude corresponds to the radius in the plot and the longitude 
    to the azimuth.}
    \label{gal}
  \end{figure}

\begin{figure}[t!]
    \centering
    \includegraphics[width=9cm,angle=0]{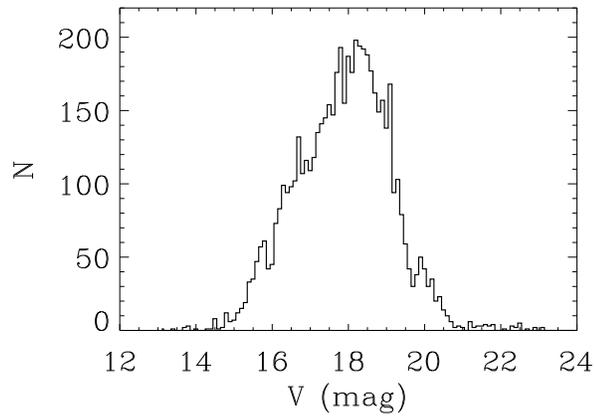}
    \caption{Distribution of brightness for the EDR stars. 
    $V$ magnitudes were
    not directly measured, but inferred from the SDSS $ugriz$ fluxes following
    Zhao \& Newberg (2002).}
    \label{nv}
  \end{figure}

The SDSS begun standard operations in April 2000. The Early Data
Release (EDR; Stoughton et al. 2002) was made public on June 5, 2001. It
consists of  462 square degrees of imaging data and 
54008 spectra. The data were acquired in three regions,
two of them following the celestial equator in the southern and
northern Galactic skies, and a third which overlaps with 
the SIRTF First Look Survey (Storrie-Lombardi et al. 2001). We 
have selected the spectra that were finally identified as {\it stellar}.
Our sample includes 5604 objects, but we were only able to identify 
Balmer lines in 4714, whose distribution in Galactic coordinates
is shown in Fig. \ref{gal}. The brightness distribution of the sample 
ranges between $V = $ 14 and 21, and it is depicted in Fig. \ref{nv}.

\begin{figure}[h!]
    \centering
    \includegraphics[width=8cm,angle=90]{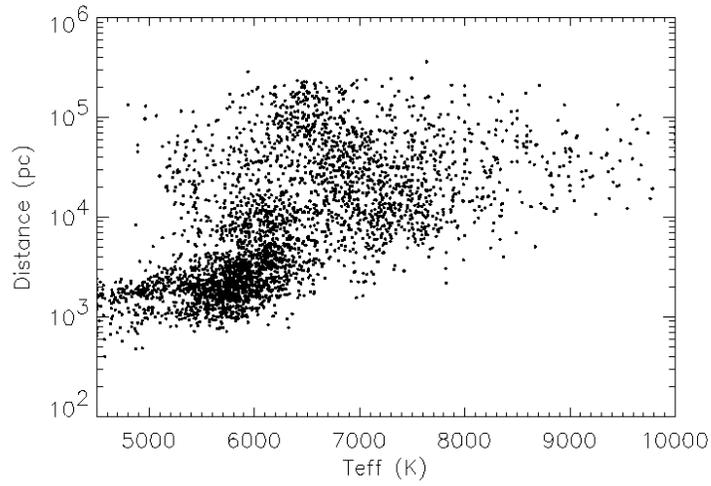}
    \caption{Position of the EDR stars in the $T_{\rm eff} -$ distance plane.}
    \label{teffd}
  \end{figure}

It is interesting to compare the distances found for the sample as a
function of the stellar $T_{\rm eff}$. In Fig. \ref{teffd}, 
dwarf stars define a tilted band
that marks the minimum distance at a given $T_{\rm eff}$. The K-type
dwarfs in the sample
are located at distances of 1--2 Kpc. Warmer stars on the main
sequence are more distant, with an obvious drop in density at 
$T_{\rm eff} \simeq 6500 $ K. Some areas in Fig. \ref{teffd} 
are  underpopulated.
Subgiants  cause the overdensity 
in a band nearly perpendicular 
to the main-sequence that crosses it at about $T_{\rm eff} = 7400$ K. 
As stars evolve off the main-sequence, they become cooler but more luminous, 
and can be seen at larger distances. A second  band parallel to the first 
is weakly apparent intersecting the main-sequence 
at a $T_{\rm eff}$ of  $\sim 6000-6300$ K. Interpreting these
features as 'turn-offs', this diagram  marks two preferred ages for the sample.

\vspace{1cm}

  \begin{figure}[h!]
    \centering
    \includegraphics[width=8cm,angle=90]{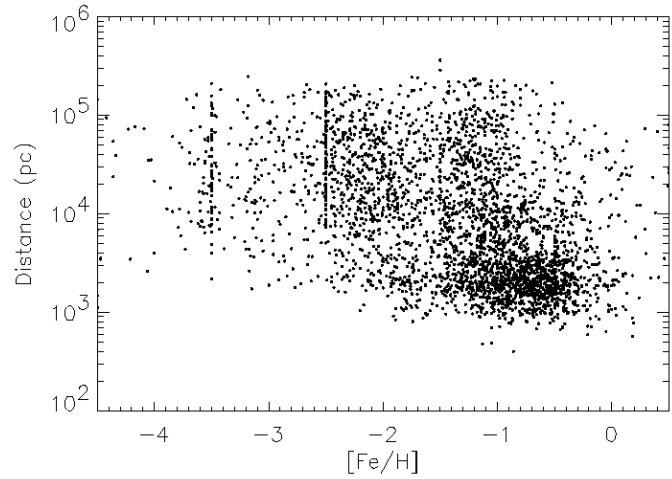}
    \caption{Position of the EDR stars in the [Fe/H] -- distance plane.}
    \label{fehd}
  \end{figure}  

The limiting magnitude for the sample of EDR stellar 
spectra is $V \simeq 20-21$.
G-type giant stars with $M_V \sim 1$  allow us to reach out to distances
as far as $\sim 50$ Kpc, but supergiants with $M_V \sim -4$ extend our scope 
about ten times farther. The sharp drop in density of stars for distances 
larger than  about 200 Kpc may or may not be a selection effect.

Another interesting projection of the data is on the plane [Fe/H] vs.
distance (see Fig. \ref{fehd}).  
The most metal-poor stars ([Fe/H]$< -3$) appear only 
 at distances larger than 3 Kpc. It is apparent in the Figure that
many stars clump at {\it small} distances ($\le 4$ Kpc) in the metallicity
range $-1.2<$[Fe/H]$< -0.4$. It is very tempting to identify this 
population with the thick disk (see, e.g., Gilmore \& Reid 1983; Reddy et al.
2003). The concentration of stars at exactly the grid nodes in [Fe/H] 
($-3.5, -2.5, \dots$)
is an artifact of the search algorithm that we do not understand yet.

  \begin{figure}
    \centering
    \includegraphics[width=8cm,angle=90]{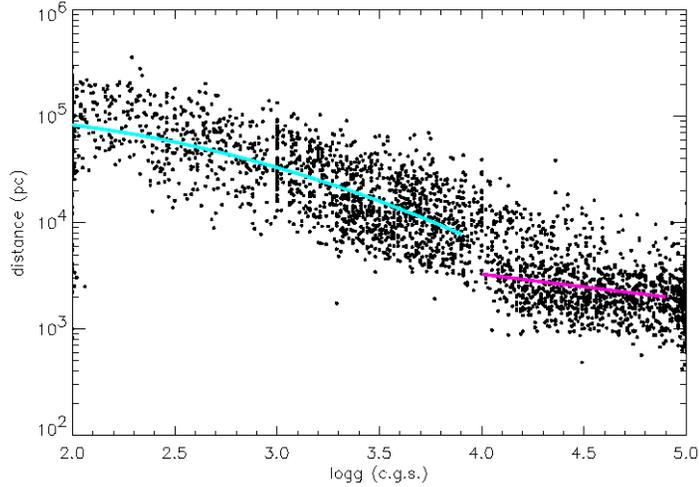}
    \caption{The graph shows which stars cover which range of distances.
    Two components  have been fitted by least-squares 
    to the data. These
     can be associated with the 
    halo (blue), and thick disk (red; but also contaminated with halo
    stars).}
    \label{loggd}
  \end{figure}  
  
  \begin{figure}
    \centering
    \includegraphics[width=10cm,angle=0]{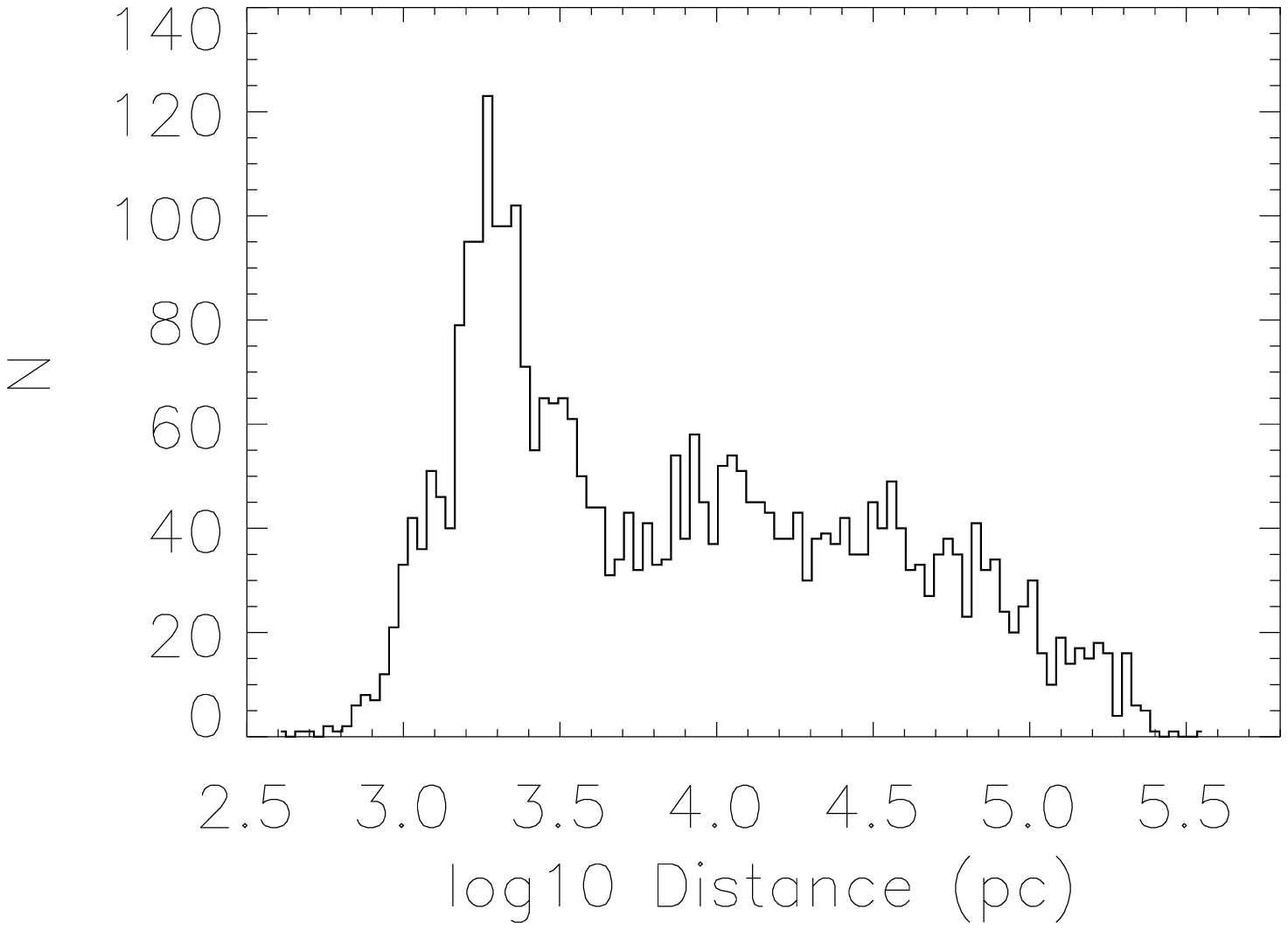}
    \includegraphics[width=10cm,angle=0]{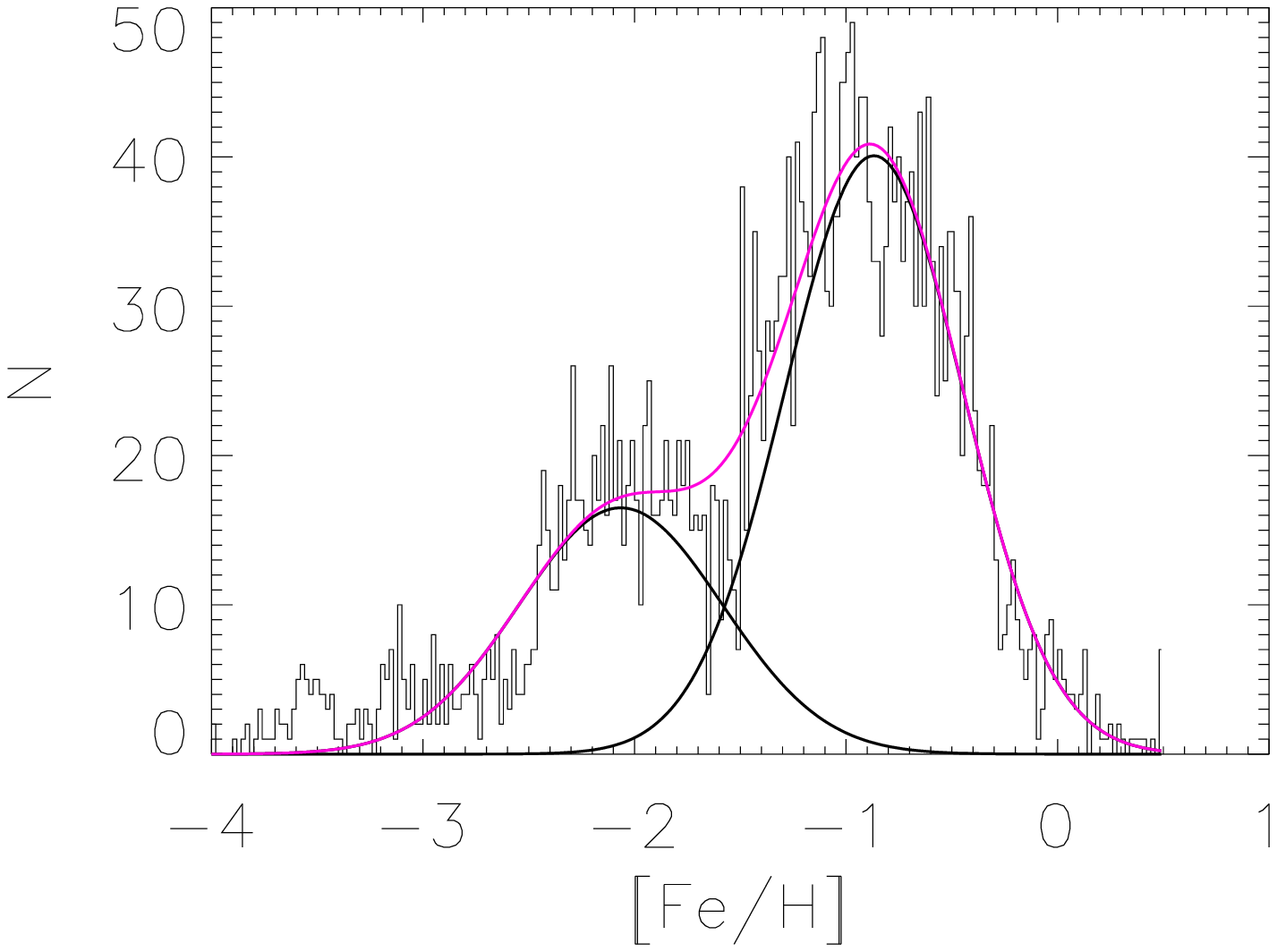}
    \caption{Histograms of the number of stars analyzed in the EDR sample
    as a function of distance (upper panel) and [Fe/H] (lower). In the lower
    panel, a two-Gaussian model has been fitted (red) curve. The individual
    components of the model are also shown in black.}
    \label{hist}
  \end{figure}

Fig. \ref{loggd} shows 
the range of distances that each luminosity class  {\it covers}. In this Figure, 
the thick disk and halo populations are clearly separated. The thick
disk stars show a 
density distribution  that rapidly falls beyond 2 Kpc. The halo star 
counts, however, decline slowly with distance. Our next challenge is
correcting the
involved selection effects to study the true density law of the halo.
  
The distribution of stars in Fig. \ref{fehd} can be collapsed 
on one axis at a time, as shown in Fig. \ref{hist}. 
In the upper panel, the clump
of stars that we identify with the thick disk becomes even more obvious. 
The distribution of brightness for the EDR stars peaks at $V \sim 18$ mag,
which corresponds to distances of $\sim 10^{4.4}$ pc for giants and
$\sim 10^{5.4}$ pc for supergiants. The continuous shape  of the number density
of stars between $10^4$ and $10^{5.5}$ pc strongly suggests that 
the observed
density decline is mainly driven by the reality in the Galactic stellar halo.

K, G, and F-type dwarfs in the EDR allow us to cover the range $\sim 10^3-10^4$ pc,
and therefore they essentially are the thick disk population in our sample
(see Fig. 1.6).
Nearby giants and supergiants brighter than $V \sim 14$ mag are rejected
by the selection algorithm to avoid saturating the detector. They 
can only cover distances larger than $10^{3.6}$ pc. Together, these selection
effects provide a simple explanation for the decrease in number density of
stars at $10^{3.7-3.8}$ pc. In fact, the star counts at $10^{3.5}$ pc seem to
recover the trend apparent at distances larger than 10 Kpc.

The lower panel of Fig. \ref{hist} reveals that the metallicity distribution 
can be approximately modeled with only two Gaussian components. A first
component, or thick-disk, centered at [Fe/H] $= -0.9$ ($\sigma = 0.4)$ dex,
and a second-component, or stellar halo, centered at [Fe/H] $= -2.1 $
($\sigma = 0.5$) dex. With the selected bin size (0.02 dex), the 
artificial overdensities at the grid nodes 
are easy to spot and have been excluded from this figure.

  \begin{figure}[h!]
    \centering
    \includegraphics[width=8cm,angle=90]{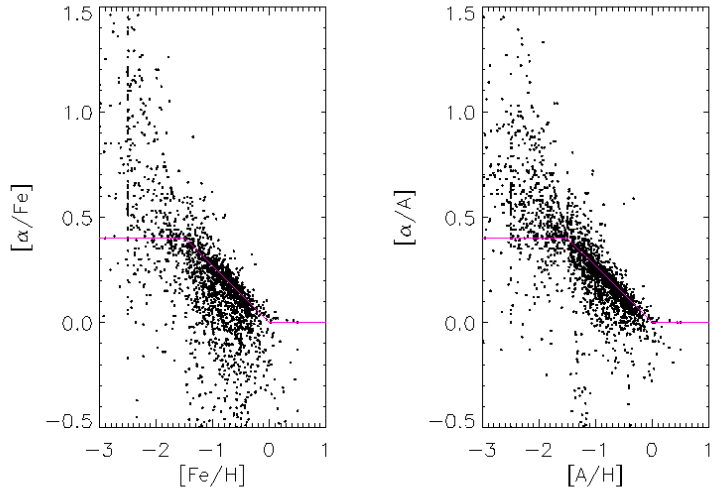}
    \caption{In an attempt to estimate the abundance ratio between the
    $\alpha$ elements (Mg and Ca) and Fe, we vary the  weights in our
    fitting procedure to favor some lines over others as metallicity
    indicators. When all lines are considered we derive [A/H], and when
    bias the analysis for, and against, lines of Mg and Ca, we obtain
    [$\alpha$/H] and [Fe/H], respectively. The red line corresponds to the
    assumption adopted for calculating the synthetic spectra, as given in
    Eq. \ref{law}.}
    \label{alpha}
  \end{figure}  
  
The weights $W_i$ in Eq. \ref{fit} are about 500 times larger for
any of the SDSS photometric indices than for any given pixel in the spectra.
The weights for the spectra are heavily biased towards lines, which carry
most of the information on the parameters we are interested in. 
The relevant lines in the optical spectrum of a metal-poor star, 
at the resolution and $S/N$ we are dealing with, 
are only a few: 
Balmer lines, Ca II H and K,
the Ca II IR triplet, the Na D lines, the Mg I b triplet, and a number of 
strong lines of the iron-peak elements (mainly Fe I). By adjusting the
weights, one has the capability of using some groups of lines and disregarding
others, biasing the results.

In Fig. \ref{alpha}, we have renamed the 
metallicities derived in the standard
case of using all possible lines (previously referred to as [Fe/H]) as [A/H].
The results of two new runs where the lines of Ca and Mg were given enhanced
weights, while lines of the iron-peak elements were disregarded, and viceversa,
are labeled as [$\alpha$/H] and [Fe/H], respectively. For the most metal-poor
stars, in particular for the warmer stars, most metal lines are too weak for
 detection in the EDR spectra, and  Ca II K remains as the only  reliable
metallicity indicator. In that regime, we expect [$\alpha$/Fe] 
to increase quickly, even exceeding unity, and [$\alpha/$A] $\rightarrow 1$.
This may be happening  for the lowest values of [Fe/H] or [A/H] 
in Fig. \ref{alpha}. However, both [$\alpha$/Fe] and [$\alpha$/A] may still
be reliable at [Fe/H] $\simeq$ [A/H] $\simeq -2$, where a clear discrepancy
is shown between measurements and our assumptions about the enhancement of
$\alpha$ elements in Eq. \ref{law} (shown as the red curve in Fig. \ref{alpha}).
The agreement is, however, fair in the interval $-1.5<$ [Fe/H] $\simeq$ 
[A/H] $<0$.

\section{Conclusions}

The spectra of Galactic 
stars acquired in the course of the Sloan Digital Sky Survey (SDSS), 
although collected mainly for calibration purposes,  
constitute an unprecedented database to study some of the oldest 
stellar populations in the Milky Way. 
The stellar atmospheric parameters and the interstellar reddening are
directly disentangled with reasonable accuracy from the spectra and 
broad-band 
photometry through standard spectral analysis techniques. 
Chemical abundances
for selected elements can also be extracted. 
Given the volume of data -- SDSS will probably obtain spectra
for $\sim 10^5$ stars --   the analysis requires automated procedures
that we implement through a pre-calculated grid of model fluxes coupled
to an optimization algorithm.
Stellar evolution theory 
allows us to constrain interesting stellar parameters such as masses, radii,
and ages, as well as to estimate distances, once the atmospheric parameters
have been derived.

A preliminary analysis of nearly 5000 spectra in the Early Data Release (EDR)
shows most SDSS dwarf stars belong to the thick disk of the Milky Way. 
This population is consistent with a 
scale height of  $1-2$ Kpc, in agreement with previous results. 
Giants and supergiants trace the Galactic halo up to 200 Kpc 
from the plane of the 
disk. The separation between thick disk and halo is evident in the range
of distances they occupy, and their chemical abundances. 
We also expect these two populations
to be distinct in soon-to-be explored ages and kinematics.

{\it  Funding for the creation and distribution of the SDSS 
Archive has been provided by the Alfred P. Sloan Foundation, 
the Participating Institutions, the National Aeronautics and 
Space Administration, the National Science Foundation, the U.S. 
Department of Energy, the Japanese Monbukagakusho, and the Max 
Planck Society. The SDSS Web site is http://www.sdss.org/.

    The SDSS is managed by the Astrophysical Research Consortium 
    (ARC) for the Participating Institutions. The Participating 
    Institutions are The University of Chicago, Fermilab, the 
    Institute for Advanced Study, the Japan Participation Group, 
    The Johns Hopkins University, Los Alamos National Laboratory, 
    the Max-Planck-Institute for Astronomy (MPIA), the Max-Planck-Institute 
    for Astrophysics (MPA), New Mexico State University, University of 
    Pittsburgh, Princeton University, the United States Naval Observatory, 
    and the University of Washington.
    
    We gratefully acknowledge NASA (grants ADP 02-0032-0106 and 
    LTSA 02-0017-0093) and NSF support 
    (grants AST 00-98508, AST 00-98549, and AST 00-86321). 
    We are thankful to David L. Carroll and 
     Alan J. Miller for making their optimization codes publicly available.
    This  research has made use of NASA's ADS.
    We  thank Bengt Gustafsson and David Lambert for interesting discussions
    and encouragement, and congratulate the organizers for an impeccable job.
}


\begin{thereferences}{}

\bibitem{}
Alongi, M., Bertelli, 
G., Bressan, A., Chiosi, C., Fagotto, F., Greggio, L., \& Nasi, E.\ 1993, 
\aas, 97, 851 

\bibitem{}
Alonso, A., Arribas, S., \& Mart\'{\i}nez-Roger, C. 1996, \aa, 313, 873

\bibitem{}
Alonso, A., Arribas, S., \& Mart\'{\i}nez-Roger, C. 1999, \aas, 140, 261

\bibitem{}
Beers, T.~C., Kage, J.~A., Preston, G.~W., \& Shectman, S.~A.\ 1990, \aj, 
100, 849 

\bibitem{}
Beers, 
T.~C., Preston, G.~W., \& Shectman, S.~A.\ 1992, \aj, 103, 1987 

\bibitem{}
Beers, T. C., Rossi, S., Norris, J. E., Ryan, S. G., \& Shefler, T. 1999,
	\aj, 117, 981
	
\bibitem{}
Bertelli, G., Bressan, 
A., Chiosi, C., Fagotto, F., \& Nasi, E.\ 1994, \aas, 106, 275 

\bibitem{}
Blackwell, D.~E., Shallis, M.~J., \& Selby, M.~J.\ 1979, \mnras, 188, 847 

\bibitem{}
Bressan, A., Fagotto, F., Bertelli, G., \& Chiosi, C.\ 1993, \aas, 100, 
647 
	
\bibitem{}
Carroll, D. L. 1999, FORTRAN Genetic Algorithm Driver, {\tt http://cuaerospace.com/carroll/ga.html}
	
\bibitem{}
Eisenstein, D. et al. 2001, \aj, 122, 2267

\bibitem{}
Fagotto, F., Bressan, A., Bertelli, G., \& Chiosi, C.\ 1994, \aas, 104, 
365 

\bibitem{}
Fitzpatrick, E. L. 1999, \pasp, 111, 63

\bibitem{}
Gilmore, G., \& Reid, N. 1983, \mnras, 202, 1025

\bibitem{}
Hubeny, I. 1998, Comp. Phys. Comm., 52, 103

\bibitem{}
Hubeny, I., Hummer, D. G., \& Lanz, T. 1994, \aa, 282, 151

\bibitem{}
Hubeny, I., \& Lanz T. 2000, Synspec: -- A User's Guide, available from {\tt
	http://tlusty.gsfc.nasa.gov}
	

\bibitem{}
Kurucz, R. L. 1993, ATLAS9 Stellar Atmosphere Programs and 2 km/s grid. Kurucz 
	CD-ROM No. 13, Cambridge, Mass.: SAO
	
\bibitem{}
Nelder, J., \& Mead, R. 1965, Computer Journal, 7, 308	

\bibitem{}
Reddy, B. E., Tomkin, J. Lambert, D. L., \& Allende Prieto, C. 2003, \mnras, 
340, 304

\bibitem{}
Richards, G. et al. 2002, \aj, 123, 2945

\bibitem{}
Storrie-Lombardi, L. J. 2001, The FIRST Look Survey Team 2001, 
Deep Fields, Proceedings of the ESO/ECF/STScI Workshop held at Garching, 
	Germany 9-12 October 2000, S. Cristiani, A. Renzini, 
	R. E. Williams, eds.,  Springer, p. 168. 

\bibitem{}
Stoughton, C. et al. 2002, \aj, 123, 485

\bibitem{}
Strauss, M.  et al. 2002, \aj, 124, 1810

\bibitem{}
Strauss, M., \& Gunn, J. E. 2001, Technical Note available from
{\tt http://archive.stsci.edu/sdss/documents/response.dat}

\bibitem{}
York, D.~G.~et al.\ 2000, \aj, 120, 1579 

\bibitem{}
Zhao, C. S., \& Newberg, H. J. 2002, technical note (private communication)
 


\end{thereferences}
\end{document}